\documentstyle[amstex,epsfig,floats,twocolumn,prl,aps,tabularx,hhline]{revtex}

\newcommand{\channel}{\mbox{$n\pi^{+}\pi^{-}\pi^{\circ}\eta$}}
\newcommand{\backchan}{\mbox{$n\pi^{+}\pi^{-}\pi^{\circ}\pi^{\circ} $}}

\begin{document}
\wideabs{
\title{Observation of a New $\bbox{J^{PC}=1^{+-}}$ Isoscalar State \\ 
in the Reaction $\bbox{\pi^-p \rightarrow \omega \eta n}$ at 18 GeV/c}

\author{
BNL-E852 Collaboration \vspace{0.1 in} \\
P.~Eugenio,$^{3}$\cite{cmu}  
G.~S.~Adams,$^{4}$ 
T.~Adams,$^{5}$\cite{ksu}  
Z.~Bar-Yam,$^{3}$ 
J.~M.~Bishop,$^{5}$ 
V.~A.~Bodyagin,$^{2}$ 
B.~B.~Brabson,$^{6}$ 
D.~S.~Brown,$^{7}$ 
N.~M.~Cason,$^{5}$ 
S.~U.~Chung,$^{1}$ 
R.~R.~Crittenden,$^{6}$ 
J.~P.~Cummings,$^{3,4}$ 
A.~I.~Demianov,$^{2}$ 
S.~Denisov,$^{8}$ 
V.~Dorofeev,$^{8}$ 
J.~P.~Dowd,$^{3}$ 
A.~R.~Dzierba,$^{6}$ 
A.~M.~Gribushin,$^{2}$ 
J.~Gunter,$^{6}$ 
R.~W.~Hackenburg,$^{1}$ 
M.~Hayek,$^{3}$\cite{haifa} 
E.~I.~Ivanov,$^{5}$ 
I.~Kachaev,$^{8}$ 
W.~Kern,$^{3}$ 
E.~King,$^{3}$ 
O.~L.~Kodolova,$^{2}$ 
V.~L.~Korotkikh,$^{2}$ 
M.~A.~Kostin,$^{2}$ 
J.~Kuhn,$^{4}$ 
R.~Lindenbusch,$^{6}$ 
V.~Lipaev,$^{8}$ 
J.~M.~LoSecco,$^{5}$ 
J.~J.~Manak,$^{5}$\cite{jlab} 
J.~Napolitano,$^{4}$ 
M.~Nozar,$^{4}$ 
C.~Olchanski,$^{1}$ 
A.~I.~Ostrovidov,$^{1,2,3}$ 
T.~K.~Pedlar,$^{7}$ 
A.~Popov,$^{8}$ 
D.~R.~Rust,$^{6}$ 
D.~Ryabchikov,$^{8}$  
A.~H.~Sanjari,$^{5}$ 
L.~I.~Sarycheva,$^{2}$ 
E.~Scott,$^{6}$ 
K.~K.~Seth,$^{7}$
N.~Shenhav,$^{3}$\cite{haifa}
W.~D.~Shephard,$^{5}$ 
N.~B.~Sinev,$^{2}$ 
J.~A.~Smith,$^{4}$ 
P.~T.~Smith,$^{6}$ 
D.~L.~Stienike,$^{5}$ 
T.~Sulanke,$^{6}$ 
S.~A.~Taegar,$^{5}$\cite{arizona}  
S.~Teige,$^{6}$ 
D.~R.~Thompson,$^{5}$ 
I.~N.~Vardanyan,$^{2}$ 
D.~P.~Weygand,$^{1}$\cite{jlab} 
D.~White,$^{4}$ 
H.~J.~Willutzki,$^{1}$ 
J.~Wise,$^{7}$ 
M.~Witkowski,$^{4}$ 
A.~A.~Yershov,$^{2}$ 
D.~Zhao$^{7}$  \vspace{0.1 in} \\
}

\address{
$^{1}$Brookhaven National Laboratory, Upton, New York 11973 \\
$^{2}$Nuclear Physics Institute, Moscow State University, Moscow, 
Russia 119899 \\
$^{3}$Department of Physics, University of Massachusetts Dartmouth, 
North Dartmouth, Massachusetts 02747 \\
$^{4}$Department of Physics, Rensselaer Polytechnic Institute, Troy, 
New York 12180 \\
$^{5}$Department of Physics, University of Notre Dame, Notre Dame, 
Indiana 46556 \\
$^{6}$Department of Physics, Indiana University, Bloomington, Indiana 47405 \\
$^{7}$Department of Physics, Northwestern University, Evanston, 
Illinois 60208 \\
$^{8}$Institute for High Energy Physics, Protvino, Russia 142284
}

\date{\today}
\maketitle

\begin{abstract}

Results are presented on a partial wave analysis of the $\omega\eta$ final state produced in $\pi^-p$ interactions at 18 GeV/{\it{c}} where $\omega \rightarrow \pi^+\pi^-\pi^0$, $\pi^0 \rightarrow 2\gamma$ , and $\eta \rightarrow 2\gamma$.  We observe the previously unreported decay mode  $\omega(1650) \rightarrow \omega\eta$  and  a new $1^{+-}$ meson state $h_1(1595)$ with a mass $M=1594(\pm15)\binom{+10}{-60}$ MeV/{\it{c}$^2$} and a width $\Gamma=384(\pm60)\binom{ +70}{-100} $ MeV/{\it{c}$^2$}. The  $h_1(1595)$ state exhibits resonant-like  phase motion relative to the $\omega(1650)$.

\end{abstract}

}

\narrowtext

\section*{ Introduction}

\label{sec:e852_introduction}

\hspace{2em} Studies of meson spectra via strong decays of hadrons provide insight
 regarding QCD at the confinement scale.  These studies have led to phenomenological 
models  such as the constituent quark model.  However, QCD demands a much 
richer spectrum of meson states which includes extra states such as 
hybrids($q\bar{q}g$), multiquarks ($q\bar{q}q\bar{q}$), and glueballs ($gg$ or $ggg$).
  Experiment E852 at Brookhaven National Laboratory is an experiment in meson 
spectroscopy configured to detect  both neutral and charged final meson states of $\pi^-p$ 
collisions in a search for meson states beyond those compatible with the constituent quark model. 

\begin{figure}[htbp] 
\centering
\leavevmode
\centerline{
\epsfig{file=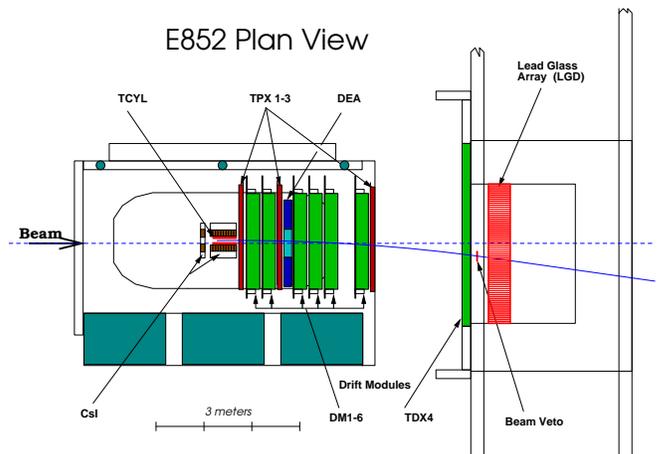,height=6cm,angle=0}}
\caption[]{The plan view of E852 Apparatus located at the Multi-Particle Spectrometer of Brookhaven's Alternating Gradient Synchrotron.}
\label{fig:E852plan}
\end{figure}

 The apparatus was located at the Multi-Particle Spectrometer  (MPS) of 
Brookhaven's Alternating Gradient Synchrotron (AGS).  The AGS delivered an 
18 GeV/{\it{c}} $\pi^-$ beam to a fixed liquid hydrogen target at the MPS.    The MPS 
facility was augmented with additional detectors designed specifically for E852 which consisted  of 3 integral regions: target, charged tracking, and downstream 
regions(see Figure \ref{fig:E852plan}).

 The target region was located in the middle of the MPS dipole magnet(1 Tesla) and contained the 
following elements: a 30.5 cm long liquid hydrogen target, a four--layer cylindrical
 drift--proportional wire chamber\cite{tcyl}, and a 198 block CsI(Ti) barrel veto detector\cite{csi}. 

 The downstream half of the MPS magnet housed the main components of the charged
 tracking region.  It consisted of 3 proportional wire chambers (TPX1--3) and 6 drift 
chamber modules(DM1--6) each with seven--layers.  In addition to the tracking 
chambers, there were two scintillation veto counters (CPVB and CPVC) and a window-frame
 lead-scintillator sandwich veto detector (DEA).  

The downstream region contained a 3000 element lead--glass calorimeter (LGD)\cite{lgd} for 
detecting and measuring the energy of the forward going gammas and a large drift 
chamber (TDX4) located directly in front of the LGD for tagging charged particles 
entering the LGD. More detailed descriptions and general features of the E852 apparatus, data acquisition, 
event reconstruction and selection are given in References \cite{mythesis,a0,etapi}.

The $\omega \eta$ final state is of considerable interest because it has been virtually unexplored, and since it has not been observed in the decay modes of known mesons.   Also, Close and Page\cite{closeNpage}, using an extension of the flux-tube model of Isgur and Paton\cite{isgurNpaton}, suggest that the isoscalar $J^{PC}=1^{--}$ $q\bar{q}g$ hybrid should decay dominantly to $\rho\pi$ and $\omega\eta$. 

The exclusive $\omega\eta$ system has been studied by two experiments previously. 
The GAMS Collaboration observed less than 100 $\omega\eta$ events produced in 
$\pi^- p$ interactions\cite{gams} and claimed to observe a narrow (less than
 50 MeV/{\it{c}}$^2$ wide) structure at 1650  MeV/{\it{c}}$^2$.  Photoproduction of  
nearly 100 $\omega\eta$ final state events was observed by the Omega Photon 
Collaboration\cite{omega_photon}.  They reported observing a peak in the $\omega\eta$ 
spectrum at a mass of 1610 MeV/{\it{c}}$^2$ but with a width of 230 MeV/{\it{c}}$^2$. 
Both experiments suffered from having too little data and  could not perform 
a partial wave analysis.  We report here the results of a partial wave analysis of 
approximately 20000 exclusive $\omega\eta$ events.

\section*{Features of the Data}

%
%
During the 1995 AGS data run, a sample of 750 million triggers was acquired of which 
108 million were of a type designed to enrich the yield from the reaction
  $\pi^-p\rightarrow n\pi^+\pi^- 4 \gamma$. About 6 million events containing 
$\pi^+\pi^-4\gamma$ were fully reconstructed.

%
%

The data were kinematically fitted  to select events consistent with a $\channel$ hypothesis. 
To eliminate broken and incorrectly hypothesized events, kinematic fits with confidence levels
 less than $5\%$ were excluded.

 There are six ways to combine the 4 $\gamma$'s into pairs, and one problem in interpreting the results of the  kinematic fit is deciding between these $\gamma\gamma$ combinations for ambiguous solutions.   For example, twenty-nine percent of the fitted $\channel$ events were also found to fit the $\backchan$ hypothesis.  Events of this type with a confidence level of at least  $0.01\%$ for the $\backchan$ hypothesis were removed from the data sample.  A total of 113000 $\channel$ events remained for further analysis.

\begin{figure}[htbp] 
\centering
\leavevmode
\centerline{
\epsfig{file=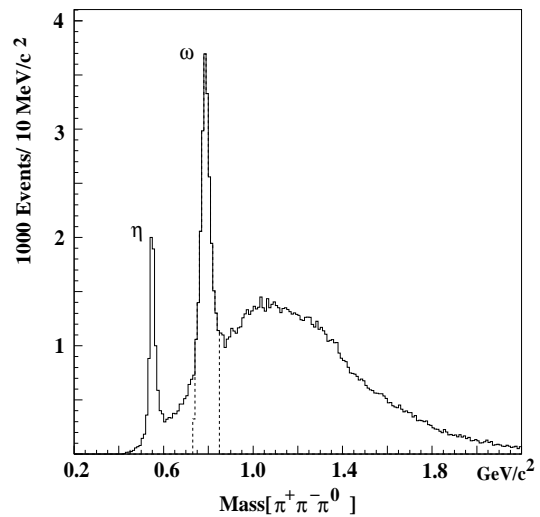,height=8cm,width=8cm}}
\caption{The $\pi^+\pi^-\pi^\circ$ effective mass distribution. Two prominent resonances are seen: the $\eta$ and the $\omega$.  Events with an invariant $\pi^+ \pi^- \pi^\circ$ mass in the region of $750 MeV$/{\it{c}$^2$} $<$ mass(3$\pi$) $<$ $830 MeV$/{\it{c}$^2$} were used to select $\omega$ events. }
\label{fig:3pi}
\end{figure}

The two prominent resonances seen in the  $\pi^+\pi^-\pi^\circ$ invariant mass distribution (see Figure \ref{fig:3pi}) are consistent with the well known $\eta$ and $\omega$.  A Gaussian plus a second order polynomial fit to these peaks results in a measured mass and width of the $\eta$ of ($550.4\pm0.2 MeV$/{\it{c}$^2$}) and  $23.4\pm0.2 MeV$/{\it{c}$^2$} and of the $\omega$ of ($787.5\pm0.3 MeV$/{\it{c}$^2$}) and $38.3\pm0.3 MeV$/{\it{c}$^2$} respectively.  The mass values are in good agreement with the PDG\cite{pdg} mass values for the $\eta$ and $\omega$, whereas the fitted widths are a measure of the experimental mass resolution of the E852 apparatus. The $\eta\rightarrow\pi^+\pi^-\pi^\circ$ events have been selected for a study of the $\eta\eta$ system which is currently in progress.  

\begin{figure}[htbp] 
\centering
\leavevmode
\centerline{
\epsfig{file=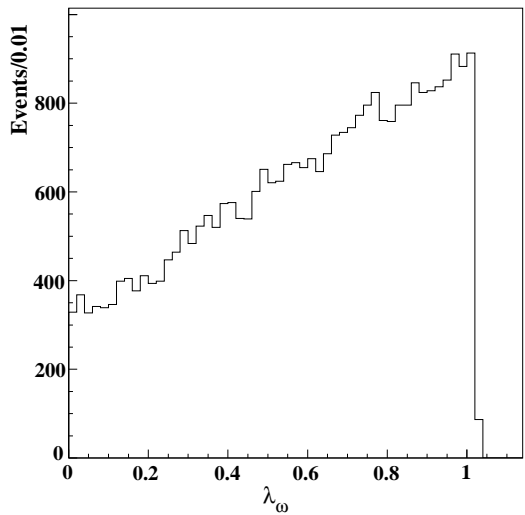,height=8cm,width=8cm}}
\caption{Distribution of events as a function of the omega decay matrix element squared, $\lambda$.  Pure $J^P=1^-$ events would exhibit a distribution which linearly increases with $\lambda$ whereas phase-space events should be flat.}
\label{fig:lambda}
\end{figure}

Events with an invariant $\pi^+\pi^-\pi^\circ$ mass in  region of $750 MeV$/{\it{c}$^2$} $<$ mass(3$\pi$) $<$ $830 MeV$/{\it{c}$^2$} were used to select $\omega$ events.  Figure \ref{fig:lambda} exhibits the $J^P=1^-$ nature of the $\omega$.   Displayed is the $\omega$ decay matrix element squared ($\lambda_\omega$) \cite{zemach}:
\[
\lambda \equiv \frac{ |\vec{\pi}^+ \times \vec{\pi}^-|^2}
{\frac{3}{4}(\frac{1}{9}M^{2}_{3\pi} - M^{2}_{\pi})^2} 
\]
where $\vec{\pi}^+$ and $\vec{\pi}^-$ are the three-momentum vectors of the the $\pi^+$ 
and $\pi^-$ in the $3\pi$ rest-frame.
The $3\pi$ decay amplitudes are constructed from the $\pi$ momentum vectors, and  due to the overall negative intrinsic parity of $3\pi$, a $J^P=1^-$ decay amplitude has to be built out of a pseudovector combination $\vec{q}$ \cite{zemach}:
\begin{equation}
\vec{q} = \vec{\pi}^+ \times \vec{\pi}^- = \vec{\pi}^- \times \vec{\pi}^0 = \vec{\pi}^0 \times \vec{\pi}^+
\end{equation}
The $\lambda$ distribution should rise linearly for omega events, whereas it would be constant for $3\pi$ events distributed according to phase-space.  The $3\pi$ background under the $\omega$ is considerable, but the  decay information of the $\omega$  can be used to weight $\omega$ events more highly than non-omega events (see the section on Background Study).

%
%
In  Figure \ref{fig:omegaeta}, the $\omega\eta$ invariant mass distribution of 19530 $\omega\eta$ final state events is shown using 40 MeV/{\it{c}$^2$} mass bins.   The invariant mass distribution rises rapidly near threshold, then increases at a slower rate to a maximum near 1600 MeV/{\it{c}$^2$}.  But, in general the mass distribution shows no clear evidence for structures indicating the presence of resonant states.

\section*{ Partial Wave Analysis}

%
%

A partial wave analysis (PWA) of the data was carried out using the 
BNL PWA program.  A general description of the BNL PWA program is
given in Reference\cite{johnNdennis}. The methods used in the PWA are based on the formalism 
of the Isobar Model\cite{isobar}.  For related material on this formalism see 
References \cite{yellowreport,PWAforms}.  

\begin{figure}[htbp] 
\centering
\leavevmode
\centerline{
\epsfig{file=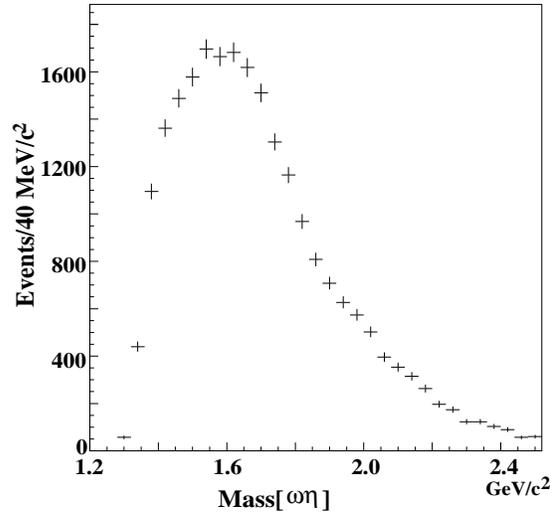,height=8cm,width=8.25cm}}
\caption{The $\omega\eta$ effective mass distribution from the reaction $\pi^- p \rightarrow 
	\omega\eta n$ (not corrected for acceptance).}
\label{fig:omegaeta}
\end{figure}

The data used in the PWA are shown in Figure \ref{fig:omegaeta}.   A PWA, taking into account both nucleon spin-flip and spin-nonflip contributions,  was performed in fifteen 50 MeV/{\it{c}$^2$} mass bins  from 1320 to 2070 MeV/{\it{c}$^2$}.

An extensive series of tests were performed to judge the stability of the PWA fit.  These include a systematic study of the PWA by  varying the:
\begin{itemize}
\item allowed waves in the fit,
\item fit starting parameters, 
\item data selection cuts, and
\item mass bin size.
\end{itemize}
In each test, the general features of the PWA fit remained unchanged.

An appropriate minimal set of partial waves needed to describe the data was determined by performing PWA fits with varying sets of partial waves.  Initially, all partial waves with $L<4$ were introduced ($L$ is the relative orbital angular momentum between the $\omega$ and the $\eta$).
Partial waves were discarded from the fit if their contributions were negligible and if the removal had little effect on the remaining features and goodness of the fit.  It was determined that a minimum set of 11 waves, which included a non-interfering  isotropic background wave, were required in order to achieve a reasonable agreement of the experimental data and the PWA fit results (see section on PWA fit quality).   A notable feature was that the exotic partial waves  for $J^{PC}=0^{--}$ and $2^{+-}$ were found not to be required in the analysis.

Table \ref{table:setA} lists the partial waves included in the final PWA fit.  Also included was an isotropic, non-omega, non-interfering background wave. The amplitudes are expressed in the reflectivity basis which takes into account parity conservation in the production process by a transformation of helicity states to eigenstates of the  production plane reflection operator\cite{chungNtrueman}.  For $\pi p$ interactions, the reflectivity coincides with the naturality of the exchange particle and amplitudes of different reflectivity $\epsilon=\pm$ do not interfere.

\setlength{\arrayrulewidth}{1pt}
\begin{table}
\caption{The minimal set of partial waves required in the PWA of the $\omega\eta$ system.}
\bigskip
\hspace{0.1 in} Partial Waves $J^{PC}\,M^\epsilon \,L$    \\ 
\centerline{
\begin{tabular}{p{4cm} p{4cm}}
 \hline{}
{\it{unnatural parity exchange}}& {\it{natural parity exchange}}\\ \hline{}
$2^{--}\,1^-P\;$   &  $1^{--}\,1^+P\;$  \\
$3^{--}\,0^-F\;$  &  $1^{+-}\,0^+S\;$  \\
   &   $1^{+-}\,0^+D\;$ \\
    &  $1^{+-}\,1^+S\;$  \\
    &  $1^{+-}\,1^+D\;$  \\
    &   $2^{--}\,0^+P\;$  \\
   &   $2^{--}\,1^+P\;$ \\
   &   $3^{--}\,1^+F\;$ \\
 \hline{}
\end{tabular}}
\label{table:setA}
\end{table}

%
\section*{  Quality of the PWA Fits}

The best test of the PWA fit is to compare the events generated according to the PWA to the data itself.  For this step, Monte Carlo(MC) events were subjected to a simulation of the E852 apparatus and weighted with the probability obtained from the PWA fit. These weighted MC events should mimic the data for successful PWA fits.  Since the fits were performed in independent mass bins, the MC events were normalized to the number of experimentally observed events in the corresponding mass bin:
\begin{equation}
\sum^{N_{MC}}_i w_i = N_{observed}.
\end{equation}

\begin{figure}[htbp] 
\centering
\leavevmode
\centerline{
\epsfig{file=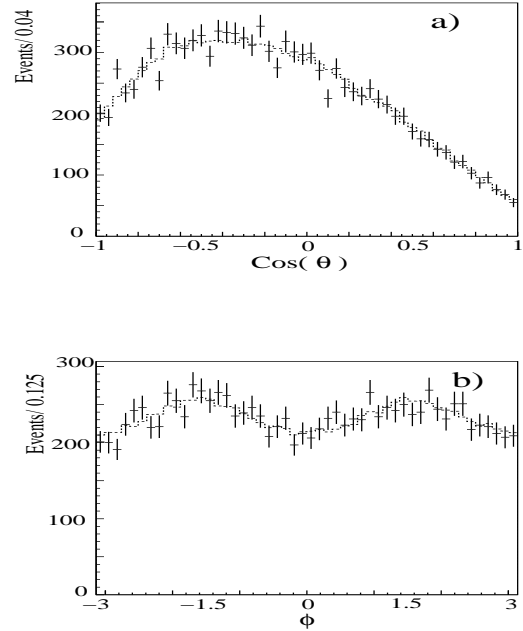,height=9cm,width=8cm}}
\caption{The quality of the fit is shown by a comparison of  angular
 distributions for the data (with error bars) and PWA fit predicted Monte Carlo 
(dashed histogram): the Gottfried-Jackson frame  a) Cos($\theta$), and b) $\phi$. }
\label{fig:angles1}
\end{figure}

\begin{figure}[htbp] 
\centering
\leavevmode
\centerline{
\epsfig{file=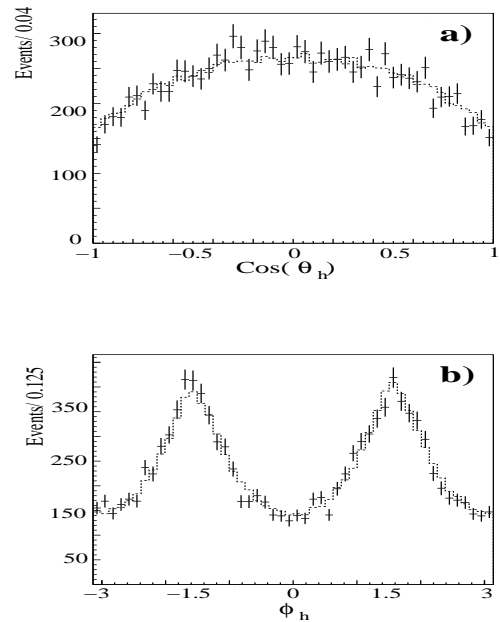,height=9cm,width=8cm}}
\caption{The quality of the fit is shown by a comparison of  angular
 distributions for the data (with error bars) and PWA fit predicted Monte Carlo 
(dashed histogram): the omega helicity frame c) Cos($\theta_h$), and d) $\phi_h$.}
\label{fig:angles2}
\end{figure}

Figures \ref{fig:angles1} and \ref{fig:angles2}  shows a comparison of the weighted MC events to the data for the: Gottfried-Jackson angles $\Omega$, and helicity angles $\Omega_h$.  The weighted MC events is in good agreement with the data.  A more stringent test compares the  angular moments, $H(l,m;L,M)$\footnote{The angular moments are the average value of the product of the two Wigner D functions $H(l,m;L,M) \equiv$ $\langle D^L_{Mm}(\Omega) *  D^l_{m0}(\Omega')\rangle$.}  . The results for this comparison (see Reference \cite{mythesis}) show good agreement between data and predicted events.  Overall, the comparisons of the distributions demonstrate that the data is very well described by the PWA fit results.

%
%
\section*{PWA Results }

\begin{figure}[htbp] 
\centering
\leavevmode
\centerline{
\epsfig{file=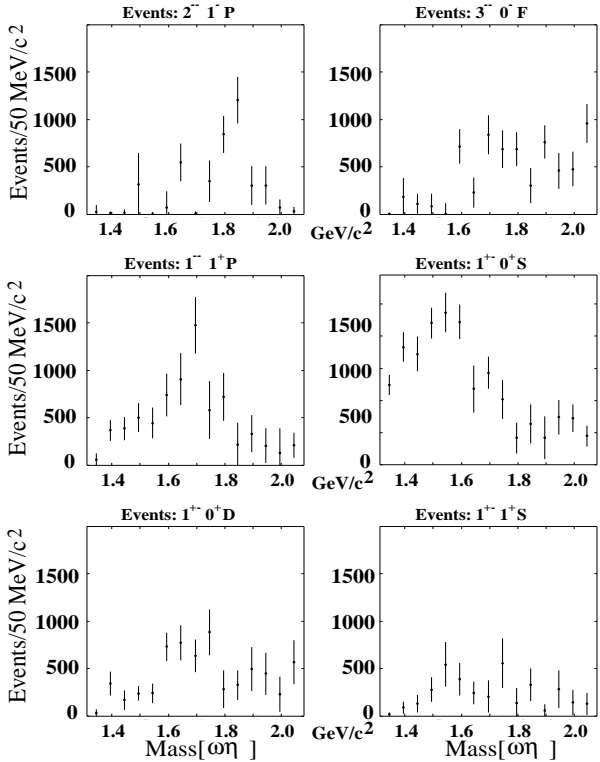 ,height=10.5cm,width=8.5cm}}
\caption{ Acceptance-corrected partial waves intensities for ($J^{PC}M^\epsilon L$): $2^{--}1^-P, 3^{--}0^-F, 1^{--}1^+P, 1^{+-}0^+S, 1^{+-}0^+D,$ and $ 1^{+-}1^+S$.
  }
\label{fig:waves1}
\end{figure}

\begin{figure}[htbp] 
\centering
\leavevmode
\centerline{
\epsfig{file=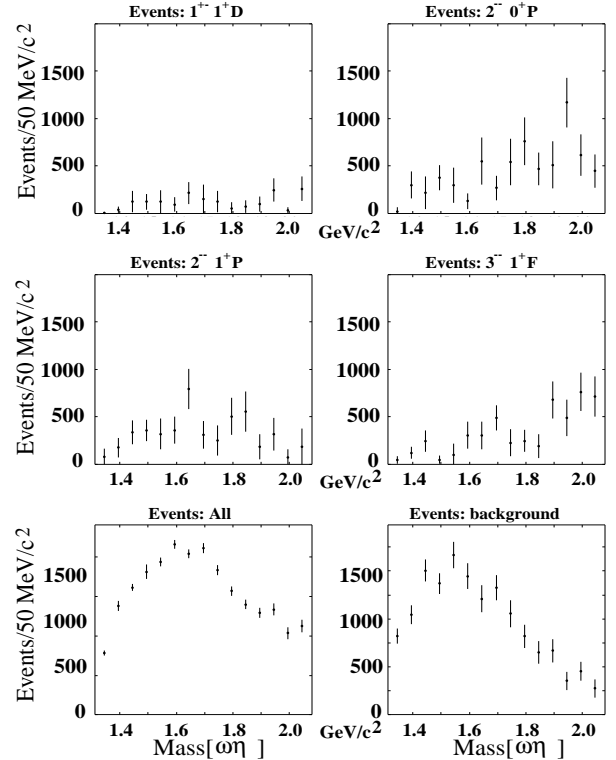 ,height=10.5cm,width=8.5cm}}
\caption{ Acceptance-corrected partial waves intensities for ($J^{PC}M^\epsilon L$): $1^{+-}1^+D, 2^{--}0^+P, 2^{--}1^+P, 3^{--}1^+F$, all waves, and background wave.
  }
\label{fig:waves2}
\end{figure}

\begin{figure}[htbp] 
\centering
\leavevmode
\centerline{
\epsfig{file=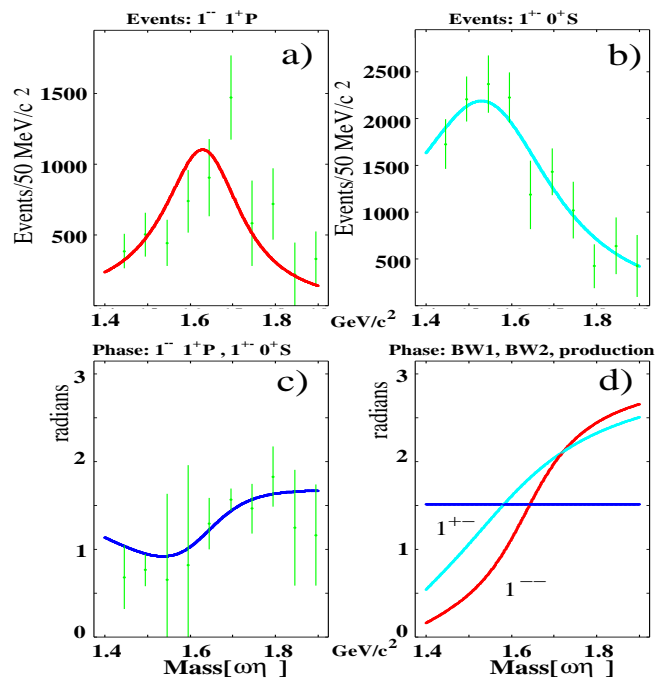,height=9.25cm,width=9.25cm}}
\caption{Mass dependent analysis of the $1^{+-}0^+S$ and $1^{--}1^+P$ partial waves 
        shows that the 2 waves are well described by two interfering 
        Breit-Wigner resonances: the $1^{--}$ state (a) fixed to the PDG parameters
	of the $\omega(1650)$ , which is the  
        first observation of $\omega(1650) \rightarrow \omega\eta$, and  a new 
        $1^{+-}$ state (b) with $mass= 1594(\pm15)\binom{+10}{-60}$ MeV/{\it{c}$^2$} and $\Gamma=384(\pm60)\binom{+70}{-100}$ MeV/{\it{c}$^2$}: c) shows the relative phase difference of $1^{+-}$ and  $1^{--}$ waves whereas d) exhibits the individual Breit-Wigner phases and an overall constant production phase.  }
\label{fig:mda}
\end{figure}

Figures \ref{fig:waves1} and  \ref{fig:waves2} show the acceptance corrected partial wave intensities.
The most significant partial waves in the analysis are the $1^{+-}0^+S$ and $1^{--}1^+P$ natural parity waves.  These two waves also exhibit relative phase motion indicative of resonant states (see Figure \ref{fig:mda}c).  Note that since the amplitudes which interfere have either purely real or purely imaginary decay factors there is an unavoidable but trivial ambiguity in the sign of the relative phase. The remaining natural parity waves are small and do not exhibit clear phase motion.  The contributions of the two unnatural parity exchange waves tend to become more important at masses higher than 1700 MeV/{\it{c}}$^2$.  Their relative phase is not well determined and exhibits large errors; therefore it cannot be reliably  used to study resonance behavior. Nonetheless, the peak in the $2^{--}1^-P$ intensity was fitted with a Breit-Wigner resonance shape giving a mass of $1830\pm8$ MeV/{\it{c}}$^2$ and a width of $86\pm31$MeV/{\it{c}}$^2$. This is interesting since the quark model predicts $J^{PC}=2^{--}$ states, yet none have been observed.  A prediction by Godfrey and Isgur suggests a mass of  1700 MeV/{\it{c}}$^2$ for the lowest lying $J^{PC}=2^{--}$ $q\bar{q}$ state\cite{godfreyNisgur}.

The only previously reported isoscalar $J^{PC}=1^{+-}$ states are the $h_1(1170)$ and $h_1(1380)$\cite{pdg}. Both of these states have masses much lower than the observed structure in the $1^{+-}0^+S$ wave.  Also, the  $h_1(1380)$, believed to be the ideal mixed $s\bar{s}$ state, is not expected to be produced in $\pi^- p$ interactions due to OZI suppression. For the isoscalar $J^{PC}=1^{--}$, there are two states listed  by the Particle Data Group (PDG) in the 1000-2000 MeV/{\it{c}$^2$} range: the $\omega(1650)$ and the $\omega(1420)$\cite{pdg}.  The $\omega(1650)$, which has a PDG mass of $1649\pm24$ MeV/{\it{c}$^2$} and a width of $220\pm35$ MeV/{\it{c}$^2$}, is in good agreement with the mass and width of the observed structure in the $1^{--}1^+P$ wave.


In order to determine if  the $1^{+-}0^+S$ and the $1^{--}1^+P$ observed structures are consistent with resonance behavior, a mass dependent analysis(MDA) of the PWA results was performed.  The input quantities included the $1^{+-}0^+S$ and  $1^{--}1^+P$ partial wave intensities and their relative phase for each mass bin.  The errors were calculated using the error matrix from the PWA fit, which takes into account correlations between the intensities and relative phases.  Relativistic Breit-Wigner forms were used to parameterize the amplitudes for the two waves.  The parameters of the MDA fit included the Breit-Wigner masses, widths, and intensities.  An additional constant parameter was included to allow for a relative constant production phase between the waves.  A series of mass dependent fits were performed for different hypotheses and for the different  phase solutions. 

The primary goal of this analysis was to understand the structure in the $1^{+-}$ wave. 
A  fit of the $1^{--}$ intensity to a single Breit-Wigner resulted in a mass of $1700(\pm20)$ MeV/{\it{c}$^2$} and a width of $250(\pm50)$ MeV/{\it{c}$^2$} for a  $\chi^2/(7 dof)$ of 0.77.  These values are in reasonable agreement with the PDG values of the $\omega(1650)$, and since the $\omega(1650)$ is rather well known  (see references \cite{omega1650}), we used the $1^{--}$ signal as a ``calibration''  by identifying 
it with the $\omega(1650)$ and fixing the $1^{--}$ Breit-Wigner parameters to the PDG values and then fitted the for the $1^{+-}$ parameters. This identification would constitute the first observation of the $\omega \eta$ decay mode of the $\omega(1650)$.  Under these assumptions, a simultaneous MDA fit  to the $1^{+-}$ intensity  and the  $1^{+-}/1^{--}$ relative phase resulted in a $\chi^2/(16 dof)$ of 0.71.  The  $1^{+-}0^+S$ mass and width resulting from this fit were  $1594(\pm15)\binom{+10}{-60}$ MeV/{\it{c}$^2$} and $384(\pm60)\binom{+70}{-100}$ MeV/{\it{c}$^2$} respectively.  The quoted errors correspond to statistical and systematic uncertainties, respectively. The systematic errors were estimated by fitting the PWA results obtained for different set of partial waves which varied from 9 to 22 waves. Figure \ref{fig:mda} displays the results of the mass dependent analysis for this fit: \ref{fig:mda}$a$, \ref{fig:mda}$b$, and \ref{fig:mda}$c$ show results of this fit overlayed on the partial wave intensities and relative phase.  Figure \ref{fig:mda}$d$ shows the absolute Breit-Wigner phases for $1^{--}$ and  $1^{+-}$ and the relative constant production phase.  On the other hand, fixing the $1^{--}$ parameters to the Breit-Wigner fitted values (mass$=1700$ MeV/{\it{c}$^2$} and width $=250$ MeV/{\it{c}$^2$}) and performing a similar MDA fit resulted in a $\chi^2/(16 dof)$ of 1.28 and a $1^{+-}0^+S$ $mass=1600(\pm20)$ MeV/{\it{c}$^2$} and $\Gamma=455(\pm80)$ MeV/{\it{c}$^2$}.

If the $1^{+-}0^+S$ object is interpreted as being caused by a single resonance, then this state, an $h_1(1595)$, does not coincide with any known states. Possible interpretations for an  $h_1(1595)$ include an $h_1$ radial excitation, an $h_1$ hybrid, or a radial-hybrid mixture.  In the Godfrey-Isgur potential model\cite{godfreyNisgur}, a mass of 1780 MeV/{\it{c}$^2$} is predicted for the $2\,^3P_0\; h_1$ radial excitation.   This is approximately 180 MeV/{\it{c}$^2$} higher than the value from our analysis.
 However this model also predicts a high mass for the $h_1(1160)$ and a high mass for the $h_1(1380)$ (1220 MeV/{\it{c}$^2$} and 1470 MeV/{\it{c}$^2$}, respectively), therefore one might expect the radial state to actually lie 100-200 MeV/{\it{c}$^2$} below the Godfrey-Isgur prediction.

A calculation by T. Barnes~\cite{ted} (also see Reference~\cite{pagecipanp2000}) using a $^3P_0$ model suggests that a 1700 MeV/{\it{c}$^2$} $h_1$ radial excitation should decay almost equally to $1^{+-}S$ and $1^{+-}D$ partial waves\footnote{The calculation is mass dependent, and one expects the $S$ to $D$ wave ratio to increase for a lower mass $h_1$ state due to angular momentum barrier effects.} which is not consistent with what we observe.  Alternatively, since the flux-tube model predicts a $J^{PC}=1^{+-}$ hybrid near 1900-2000 MeV/{\it{c}$^2$}\cite{isgurNpaton}, and exotic states have been reported at masses lower than the flux-tube model expectations, an $h_1(1595)$ would be a candidate for a hybrid or radial-hybrid mixed state.

\section*{ Background Study}
The $\lambda$ distribution can be used to estimate the background due to non-omega events under the omega signal. The omega signal-to-background ratio estimated from Figure \ref{fig:lambda} is a little larger than 1 to 1. But since the omega is a spin 1 particle, the PWA uses the omega angular decay information to weight omega events more highly than non-omega events.

\begin{figure}[htbp] 
\centering
\leavevmode
\centerline{
\epsfig{file=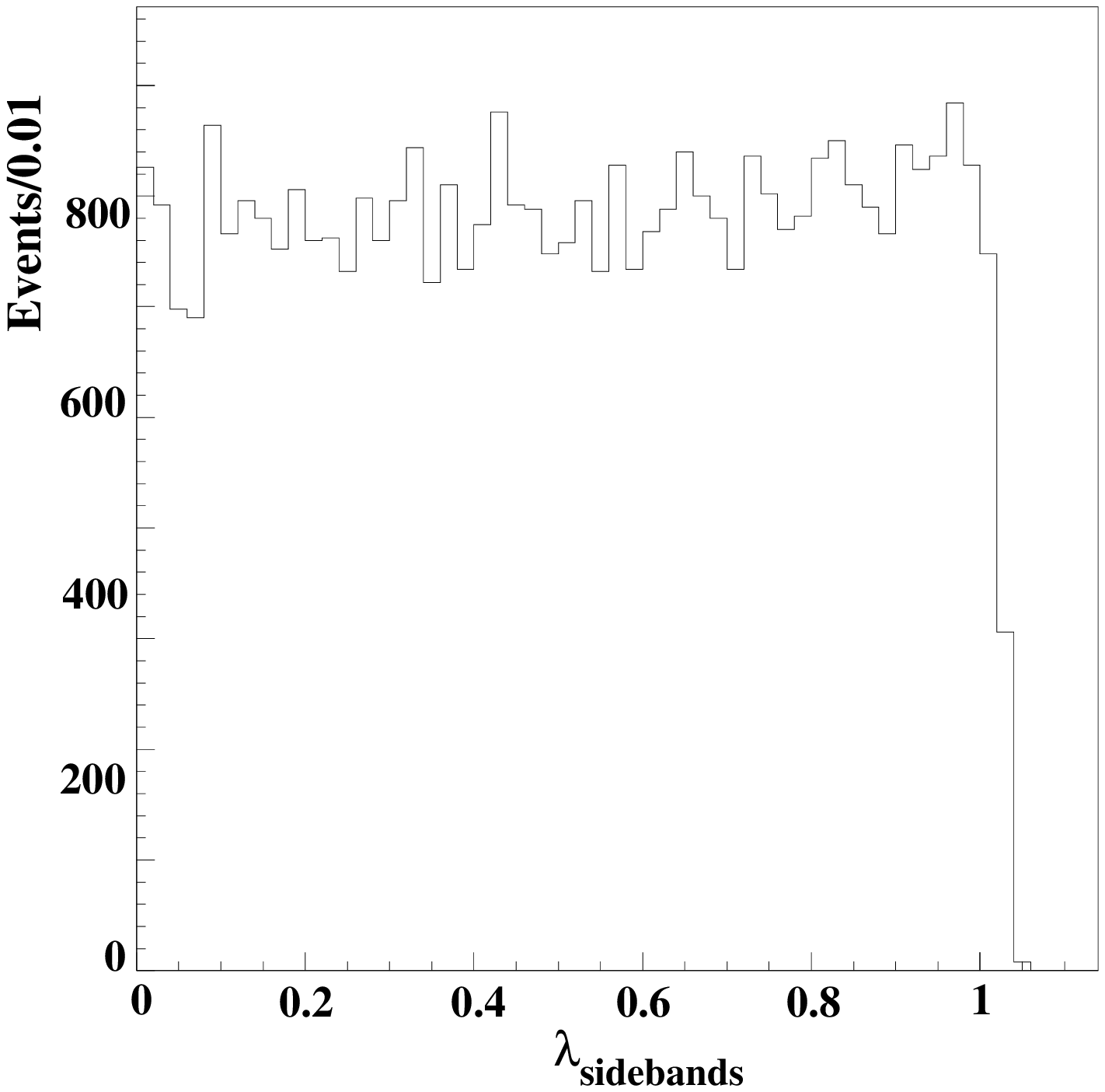,height=8cm,width=8cm}}
\caption{Distribution of the sideband events as a function of the omega decay matrix element squared, $\lambda$.  Pure $\omega$ events would exhibit a distribution which linearly increases with $\lambda$ whereas phase-space events w/ould be flat. }
\label{fig:sidebands}
\end{figure}

To better understand the background under the omega and to see if these events are responsible for the observed structure in the $1^{+-}0^+S$ and  $1^{--}1^+P$ partial waves, events were selected from the sidebands around the omega region and a partial wave analysis of these events was performed.  As expected, the $\lambda_{sidebands}$ distribution, shown in Figure \ref{fig:sidebands}, is flat and consistent with phase-space.  

\begin{figure}[htbp] 
\centering
\leavevmode
\centerline{
\epsfig{file=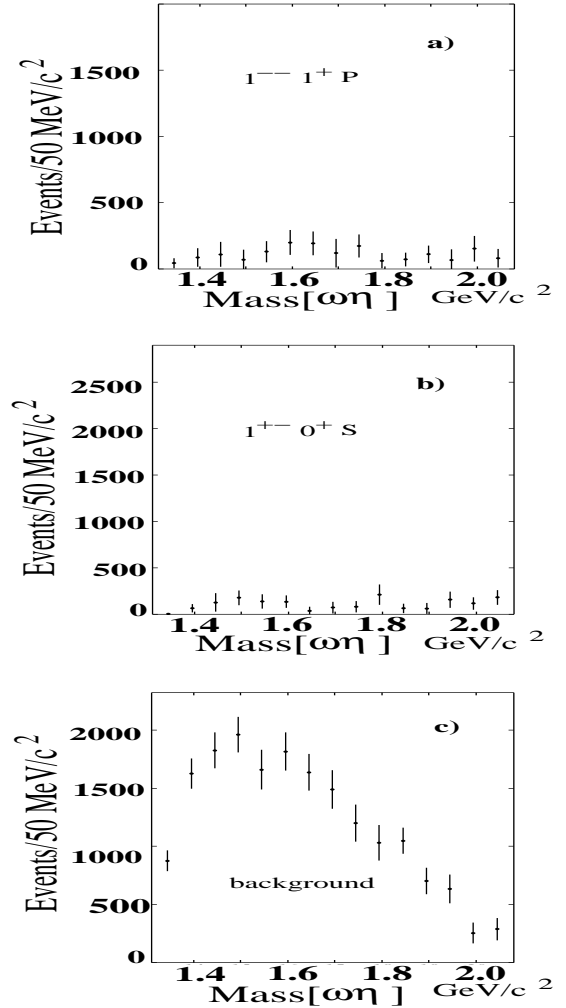,height=14cm,width=8.0cm}}
\caption{ The PWA results for the sidebands events.  Shown are the : a) $1^{--}1^+P$ intensity, b)  $1^{+-}0^+S$ intensity, and c) the background intensity. }
\label{fig:pwasides}
\end{figure}

A PWA of the sideband events was performed in the same manner as the original PWA of the data.  Decay amplitudes were calculated using the sideband events whereas the normalization integrals and the list of partial waves included in the fit were identical to those used in the original PWA.  Figure \ref{fig:pwasides} shows  results of the sideband PWA.  Note that all partial waves except the background wave have an omega decay factor in the amplitude.  This makes the background wave more favorable to describe those events which do not contain omegas. As seen in Figures \ref{fig:pwasides}a and \ref{fig:pwasides}b, very few sideband events contribute to the $1^{+-}0^+S$ and  $1^{--}1^+P$ waves demonstrating that non-omega events are unlikely to be responsible for the observed resonant-like structures.   The sideband events overwhelmingly contribute to the background intensity as shown in Figure \ref{fig:pwasides}c.  

\section*{ Conclusions}
%
%
We have collected a high statistics sample of the reaction
\[
\pi^- p \rightarrow n \pi^+ \pi^- \pi^0 \eta
\]
at 18 GeV/{\it{c}} beam momentum, and  performed a partial wave analysis on the $\omega\eta$ system where 
\[
\omega \rightarrow \pi^+\pi^-\pi^0 \;\;\;\;\;\;\;\;\; \pi^0 \rightarrow 2\gamma \;\;\;\;\;\;\;\;\; \eta \rightarrow 2\gamma.
\]

The $\omega\eta$ invariant mass distribution rises rapidly near threshold then increases slowly to a maximum near 1600 MeV/{\it{c}$^2$}.  A PWA finds the data dominated by natural parity exchange waves.  The results show 2 significant isoscalar $J^{PC}M^\epsilon L$ partial wave intensities: a $1^{+-}0^+S$ intensity peaking near 1600 MeV/{\it{c}$^2$} with a width of about 300 MeV/{\it{c}$^2$}, and a $1^{--}1^+P$ intensity peaking near 1650 MeV/{\it{c}$^2$} also with a width of about 200 MeV/{\it{c}$^2$}. 
In addition, a peak in the unnatural parity exchange wave $2^{--}1^-P$  exhibits a Breit-Wigner mass of $1830\pm8$ MeV/{\it{c}}$^2$ and  width of $86\pm31$MeV/{\it{c}}$^2$.  Unfortunately the phase of this state was not well determined and thus claim of a $2^{--}$ state at this mass requires further confirmation.  Another notable feature was that the exotic partial waves for $J^{PC}=0^{--}$ and $2^{+-}$ were found not to be required by the data.

A mass dependent analysis of $1^{+-}0^+S$ and $1^{--}0^+P$ partial waves shows that the intensities and relative phase motion are well described by two resonating states.  A  $1^{--}1^+P$ state in good agreement with the PDG values for the $\omega(1650)$, and a new  $h_1(1595)$ state with $mass=1594(\pm15)\binom{+10}{-60}$ MeV/{\it{c}$^2$} and $\Gamma=384(\pm60)\binom{+70}{-100}$ MeV/{\it{c}}$^2$.
It is  interesting to note that Close and Page \cite{closeNpage} predict that the isoscalar $J^{PC}=1^{--}$ hybrid should decay dominantly to states containing a vector meson such as $\rho\pi$ and $\omega\eta$ and not to traditional flux-tube decays such as $b_1\pi$.
Possible interpretations for an $h_1(1595)$ include an $h_1$ radial excitation, a $h_1$ hybrid, or a $h_1$ radial-hybrid mixture.

We are grateful for the support of the technical staffs of the MPS, AGS, BNL, and various collaborating institutions.  This research was supported in part by the US Department of Energy, the National Science Foundation, and the Russian State Committee for Science and Technology.


\end{document}